\documentclass[prl, twocolumn,10pt,longbibliography,superscriptaddress]{revtex4-2}
\usepackage{textcomp}
\usepackage{amsmath}
\usepackage{amsfonts}
\usepackage{calc}
\usepackage{mathrsfs}
\usepackage{amssymb}
\usepackage{amsmath}
\usepackage{array}
\usepackage{color}
\usepackage{bm}
\usepackage{graphicx}
\usepackage{hyperref}
\usepackage{booktabs}
\usepackage[scr=boondoxo,scrscaled=1.05]{mathalfa}

\renewcommand{\P}{\mathcal{P}}
\newcommand{\res}{\mathcal{R}}

\newcommand{\e}{\epsilon}
\newcommand{\Mirr}{M_{BH}}
\newcommand{\Mhatirr}{\hat M_{BH}}

\newcommand{\beq}{\begin{equation}}
\newcommand{\eeq}{\end{equation}}
\newcommand{\modefont}{\mathscr}

\begin{document}
\title{Second-order self-force calculation of the gravitational binding energy in compact binaries} 
\author{Adam Pound} 
\affiliation{School of Mathematical Sciences and STAG Research Centre, University of Southampton, Southampton, United Kingdom, SO17 1BJ}
\author{Barry Wardell}
\author{Niels Warburton}
\affiliation{School of Mathematics and Statistics, University College Dublin, Belfield, Dublin 4, Ireland}
\author{Jeremy Miller}
\affiliation{School of Mathematical Sciences and STAG Research Centre, University of Southampton, Southampton, United Kingdom, SO17 1BJ}
\date{\today}

\begin{abstract}
Self-force theory is the leading method of modeling extreme-mass-ratio inspirals (EMRIs), key sources for the gravitational-wave detector LISA. It is well known that for an accurate EMRI model, {\em second-order} self-force effects are critical, but calculations of these effects have been beset by obstacles. In this letter we present the first implementation of a complete scheme for second-order self-force computations, specialized to the case of quasicircular orbits about a Schwarzschild black hole. As a demonstration, we calculate the gravitational binding energy of these binaries.
\end{abstract}
\maketitle

Recent years have seen profound advances in our knowledge of the gravitational two-body problem. The LIGO and Virgo Collaborations' landmark detections of gravitational waves have provided our first observations of highly relativistic two-body systems in nature~\cite{LIGOdetection1,LIGO-catalog,LIGO-alerts}. These observations have been enabled by (and have validated) our theoretical models of such systems, without which the observed waves would be unintelligible. Great strides continue to be made in these models throughout the binary parameter space: in the venerable post-Newtonian (PN) theory~\cite{Damour-etal:14,Marchand-etal:17,Foffa-etal:19}, which applies for binaries with large orbital separations; in numerical relativity (NR)~\cite{Baumgarte-Shapiro:10}, which applies for small separations and comparable masses; and in post-geodesic, gravitational self-force (SF) theory~\cite{Warburton-Osburn-Evans:17,vandeMeent:17b}, which applies for disparate masses. 

By interfacing in regimes of mutual validity, these distinct models have also fruitfully informed and improved one another, often in unforeseen ways~\cite{LeTiec:14,Blanchet-etal:10b,Bini-etal:18,Damour-etal:16,LeTiec-etal:11,LeTiec-etal:12b}. Concurrently, the phenomenological effective-one-body theory~\cite{Buonanno-Damour:99} has begun to synthesize them into a single, universal model of two-body systems~\cite{Bohe-etal:16, Akcay:2018yyh, Antonelli:2019fmq}. 

However, despite these advances, modeling has remained critically limited in the small-mass-ratio, SF regime. For ground-based detectors such as LIGO, comparable-mass binaries are the dominant sources, and consequently the most effort has gone into modeling them. However, when it launches, the space-based detector LISA will observe extreme-mass-ratio inspirals (EMRIs), in which stellar-mass BHs or neutron stars spiral into supermassive BHs~\cite{Babak-etal:17}. These systems will allow us to precisely measure corrections to the test-particle, geodesic approximation for the small companion in the binary, and the intricate waveforms they emit will encode incomparably precise maps of the supermassive BHs' spacetimes~\cite{Amaro-Seoane-etal:14}. Due to their unique scientific potential, EMRIs have been a prime motivator for the development of SF theory. But SF theory is based on a perturbative expansion, limited by the order to which it is carried: the small companion acts as a source of perturbation of the central BH's spacetime, and that perturbation then exerts a SF back on the small companion, driving it away from geodesic motion~\cite{Barack:09,Poisson-Pound-Vega:11,Wardell:15,Pound:15a,Barack-Pound:18}. Accurately extracting system parameters from an EMRI waveform requires carrying this expansion to second perturbative order~\cite{Hinderer-Flanagan:08}. Yet despite tremendous progress in the SF programme over the past two decades, computations at second order have remained out of reach. 


Working at second order comes with numerous challenges. At the foundational level, the point-particle approximation for the small object fails, and only in 2012 were viable formulations of second-order SF theory derived~\cite{Pound:12a,Gralla:12,Pound:12b,Pound:17}, after some years of preparatory work~\cite{Rosenthal:06a,Rosenthal:06b,Gralla-Wald:08,Pound:10a,Detweiler:12}. At the level of concrete implementation, new effects (and new obstacles) arise on both large temporal and spatial scales~\cite{Pound:15c} and on small scales near the small object~\cite{Miller-Wardell-Pound:16}. Until now, only a  partial numerical calculation, based on an incomplete theory, has ever been performed~\cite{Lousto-Nakano:08}.

In this letter, building on progress in~\cite{Pound-Miller:14,Warburton-Wardell:14,Wardell-Warburton:15, Pound:15c, Miller-Wardell-Pound:16}, we report a milestone in binary modeling: the first implementation of a complete, concrete numerical scheme for second-order SF calculations. We specialize to quasicircular orbits around a Schwarzschild BH, and in that context, we compute the binaries' gravitational binding energy.

\emph{Self-force theory.}
We begin by expanding the binary's metric ${\sf g}_{\mu\nu}$ in powers of the mass ratio $\e:=m/M$, where $m$ is the mass of the smaller object, such that ${\sf g}_{\mu\nu}(\e) = g_{\mu\nu}+\sum_{n\geq 1}\e^n h^n_{\mu\nu}$. Here $g_{\mu\nu}$ is the metric of a Schwarzschild BH of mass $M$, and we assume that the small object is nonspinning and spherical. The equations of SF theory, through second order in $\e$, are then given by~\cite{Pound:12a,Pound:12b,Pound:17}
\begin{align}
E[\bar h^1_{\mu\nu}] &= 8\pi T_{\mu\nu},\label{EFE1}\\ 
E[\bar h^{2\res}_{\mu\nu}] &= -\delta^2 G_{\mu\nu}[h^1]-E[\bar h^{2\P}_{\mu\nu}],\label{EFE2}\\
\frac{D^2z^\alpha}{d\tau^2} &= -\frac{1}{2}\left(g^{\alpha\mu}+u^\alpha u^\mu\right)g^{\rho\delta}
	\left(g_{\mu\rho}-h^{\res}_{\mu\rho}\right)\nonumber\\
	&\quad\times\left(2h^\res_{\delta\beta;\gamma}-h^\res_{\beta\gamma;\delta}\right)\!u^\beta u^\gamma,\label{EOM}
\end{align}
subject to the Lorenz gauge condition $g^{\rho\nu}(\e \bar h^1_{\mu\nu;\rho}+\e^2 \bar h^2_{\mu\nu;\rho})={\cal O}(\e^3)$. Here an overbar denotes trace-reversal, as in $\bar h^1_{\mu\nu}:=h^1_{\mu\nu}-\frac{1}{2}g_{\mu\nu}g^{\alpha\beta}h^1_{\alpha\beta}$, and a semicolon denotes covariant differentiation compatible with $g_{\mu\nu}$. $E[\bar h_{\mu\nu}]:=-\frac{1}{2}\left(g^{\alpha\beta}\bar h_{\mu\nu;\alpha\beta}+2R_\mu{}^\alpha{}_\nu{}^\beta \bar h_{\alpha\beta}\right)$ is the linearized Einstein tensor in the Lorenz gauge, and $\delta^2 G_{\mu\nu}[h^1]$ is the term quadratic in $h^1_{\mu\nu}$ in the expansion of the Einstein tensor. At leading, linear order, in Eq.~\eqref{EFE1}, the small object is represented by a point-mass stress-energy $T_{\mu\nu}$. The point mass moves on a representative worldline $z^\mu$ governed by Eq.~\eqref{EOM}, in which $\tau$ is proper time as measured in $g_{\mu\nu}$, $u^\mu:=dz^\mu/d\tau$ is the particle's four-velocity, and $\frac{D^2z^\alpha}{d\tau^2}=u^\beta u^\alpha{}_{;\beta}$ is its covariant acceleration.

Beyond linear order, the point-particle approximation breaks down, and in Eq.~\eqref{EFE2} we instead split the physical field into $h_{\mu\nu} = h^{\P}_{\mu\nu}+h^{\res}_{\mu\nu}$. $h^{\P}_{\mu\nu}=\sum_{n\geq 1}\e^n h^{n\P}_{\mu\nu}$ is a {\em puncture} in the geometry, given in covariant form in Eqs.~(126)--(133) of~\cite{Pound-Miller:14}. It diverges on $z^\mu$, but asymptotically near $z^\mu$ it is the dominant part of the physical metric outside the small object. $h^{\res}_{\mu\nu}=\sum_{n\geq 1}\e^n h^{n\res}_{\mu\nu}$ is the {\em residual} field. It governs the motion of $z^\mu$, exerting the SF (per unit mass) on the right-hand side of Eq.~\eqref{EOM}. At first order, one can solve Eq.~\eqref{EFE1} directly for $h^{1}_{\mu\nu}$, afterward recovering $h^{1\res}_{\mu\nu}=h^{1}_{\mu\nu}-h^{1\P}_{\mu\nu}$. At second order, one must instead solve Eq.~\eqref{EFE2} directly for $h^{2\res}_{\mu\nu}$, with $h^{2\P}_{\mu\nu}$ moved to the right-hand side and treated as a source; this is necessary to cancel the nonintegrable singularity in $\delta^2 G_{\mu\nu}[h^1]$ at $z^\mu$. At some distance from $z^\mu$, $h^{n\P}_{\mu\nu}$ transitions to zero, such that beyond that distance, $h^{n\res}_{\mu\nu}$ is equal to the physical field $h^{n}_{\mu\nu}$. 

\emph{Quasicircular orbits.} We now suppose that the small object is slowly spiraling into the BH along a quasicircular trajectory. To efficiently account for the system's slow evolution, we perform a multiscale expansion of $z^\mu$ and $h^n_{\mu\nu}$, following Sec. IV of~\cite{Pound:15c}. We introduce a ``slow time" variable $\tilde t:=\e t$, where $t$ is Schwarzschild time, and we write the worldline in Schwarzschild coordinates as $z^\mu(\tilde t,\e) = \{t, r_p(\tilde t,\e),\pi/2,\phi_p(\tilde t,\e)\}$. Both the orbital radius $r_p$ and frequency $\Omega:=\frac{d\phi_p}{dt}$ evolve slowly due to dissipation, on the timescale $\tilde t\sim M$, with expansions
\begin{align}
r_p(\tilde t,\e) &= r_0(\tilde t) + \e r_1 (\tilde t) + {\cal O}(\e^2), \label{rp-expansion}\\
\Omega(\tilde t,\e) &= \Omega_0(\tilde t) + \e \Omega_1 (\tilde t) + {\cal O}(\e^2).\label{Omega-expansion}
\end{align}
The azimuthal phase, which varies on the timescale $t\sim M$, is recovered from $\Omega$ as 
\beq
\phi_p(\tilde t,\e) = \int \Omega dt = \frac{1}{\e}\int\Omega(\tilde t,\e)d\tilde t.\label{phip}
\eeq

Substituting these expansions into Eq.~\eqref{EOM} leads to a sequence of equations for $dr_n/d\tilde t$ and $\Omega_n(r_n)$. In particular, the zeroth-order frequency is $\Omega_0 = \sqrt{\frac{M}{r_0^3}}$, a slowly evolving version of the usual geodesic frequency, and its first-order correction is
\beq\label{Omega1}
\Omega_1 = -\frac{1}{2r_0f_0\Omega_0}\left[(u^t_0)^{-2}F^r_1(r_0)+3\Omega_0^2f_0r_1\right],
\eeq
where $F^\alpha_1(r_0)=\frac{1}{2}g^{\alpha\beta}\partial_\beta h^{1\res}_{\mu\nu}(r_0)u_0^\mu u_0^\nu$ is the first-order SF per unit mass, $u_0^\mu=u_0^t(1,0,0,\Omega_0)$ is the zeroth-order four-velocity, $u_0^t = 1/\sqrt{1-3M/r_0}$, and $f_0:=1-2M/r_0$. These expressions provide the instantaneous frequency as a function of orbital radius; the equations for $dr_0/d\tilde t$ and $dr_1/d\tilde t$, which will not be needed explicitly here, then determine how the frequency evolves with time. 

Still following~\cite{Pound:15c}, we now note that $T_{\mu\nu}\propto\delta[r-r_p(\tilde t,\e)]\delta(\theta-\pi/2)\delta[\phi-\phi_p(\tilde t,\e)]$ is a periodic function of $\phi_p(\tilde t,\e)$. Specifically, if we expand the angular delta functions in spherical harmonics $Y_\modefont{lm}$, then $T_\modefont{lm}\propto e^{-i\modefont{m}\phi_p(\tilde t,\e)}$. This motivates us to adopt $\phi_p:=\phi_p(\tilde t,\e)$ as our ``fast time" variable and expand $T_{\mu\nu}$, $h^{n\P}_{\mu\nu}$, and $h^{n}_{\mu\nu}$ in powers of $\e$ at fixed $\tilde t$ and $\phi_p$. By simultaneously expanding in a basis of tensor spherical harmonics, and absorbing subleading corrections in $h^1_{\mu\nu}$ into the new, multiscale $h^2_{\mu\nu}$, we obtain fields of the form
\beq\label{multiscale-h}
\bar h^n_{\mu\nu} = \sum_{\modefont{ ilm}}R^n_{\modefont{ ilm}}(r,\tilde t)e^{-i{\modefont{ m}}\phi_p}Y^{\modefont{ ilm}}_{\mu\nu},
\eeq
where $Y^{\modefont{ ilm}}_{\mu\nu}$ ($\modefont{ i}=1,\ldots,10$) are Barack-Lousto-Sago harmonics~\cite{Barack-Lousto:05,Barack-Sago:07}. Following standard multiscale methods~\cite{Kevorkian-Cole:96}, we then treat $\tilde t$ and $\phi_p$ as independent variables and rewrite the field equations~\eqref{EFE1}--\eqref{EFE2} as equations for $R^n_\modefont{ ilm}(r,\tilde t)$. Using the chain rule $\partial_t = -i\modefont{ m}\Omega + \e\partial_{\tilde t}$ and the expansion~\eqref{Omega-expansion},  and regrouping according to explicit powers of $\e$ at fixed $\tilde t$ and $\phi_p$, we obtain ordinary differential equations of the form
\begin{align}
E^0_\modefont{i l m}\![R^1]\! &=\! 8\pi T^{1}_\modefont{ ilm},\label{tildeEFE1}\\
\!E^0_\modefont{ ilm}\![R^{2\res}]\! &=\! -\delta^2G^0_\modefont{ ilm}\![R^1]\!-\!E^0_\modefont{ ilm}\![R^{2\P}]\! -\! E^1_\modefont{ ilm}\![R^1],\label{tildeEFE2} 
\end{align}
where $E^0_\modefont{ ilm}$ and $\delta^2G^0_\modefont{ ilm}$ are purely radial differential operators, in which we set $\partial_t=-i\modefont{ m}\Omega_0$, and $E^{1}_\modefont{ ilm}$ is linear in $-i\modefont{ m}\Omega_1+\partial_{\tilde t}$. Equation~\eqref{tildeEFE1} is identical to a standard frequency-domain field equation for a point mass on a circular geodesic of radius $r_0$, as given in, e.g., Eq.~(2.10) of~\cite{Wardell-Warburton:15}. The stress-energy in it is $T^1_\modefont{ ilm}\propto\delta(r-r_0)$; the subleading term in $T_{\mu\nu}$ has been accounted for through additional terms ($\propto r_1$, $\Omega_1$, or $dr_0/d\tilde t$) in $R^{2\P}_\modefont{ ilm}$. Equations analogous to Eqs.~\eqref{tildeEFE1}--\eqref{tildeEFE2} also follow for the gauge condition.

On the right-hand side of Eq.~\eqref{tildeEFE2}, we require the tensor-harmonic modes of $\bar h^{2\P}_{\mu\nu}$ and $\delta^2 G_{\mu\nu}$. The former of these we obtain from the covariant expressions for $h^{2\P}_{\mu\nu}$, following Sec.~IVA of~\cite{Wardell-Warburton:15}. The latter we obtain using the method detailed in~\cite{Miller-Wardell-Pound:16}. 

As described in~\cite{Pound:15c}, the multiscale expansion breaks down near infinity and the BH horizon due to long-term evolution effects propagating over large spatial scales. To overcome this, we introduce new, analytical expansions in those regions. Near future null infinity, we use a post-Minkowskian (PM) expansion (adapted from Blanchet and Damour~\cite{Blanchet-Damour:92}), iteratively solving the field equations with the Minkowskian retarded Green's function, specifically following Sec. V of~\cite{Pound:15c}. Near the horizon, we use an analogous technique with a retarded Green's function tailored to the local geometry (inspired by~\cite{Barack:99}). These analytical expansions then provide boundary conditions for our multiscale solution. 

\emph{Snapshot of the system and specialization to $\modefont{l}=0$.} The multiscale framework can ultimately be used to simulate complete evolutions and their emitted waveforms (to ``post-adiabatic" accuracy, in the sense of~\cite{Hinderer-Flanagan:08}). From the amplitudes $R^n_\modefont{ ilm}(\tilde t,r)$, one can construct the SF; from the SF, $r_n(\tilde t)$ and $\Omega(r_n(\tilde t))$; and from $\Omega$, finally $\phi_p(\tilde t)$. Equation~\eqref{multiscale-h} then yields the full time-domain metric perturbation over the course of the inspiral. 

However, as a first implementation, we restrict ourselves to a single value of $\tilde t$, call it $\tilde t_0$. At $\tilde t_0$, we specify that $\Omega=\Omega_0$, such that our expansion in the limit $\e\to0$ is performed at fixed orbital frequency; for a given inspiral, this choice can be made freely at any one value of $\tilde t$. $r_0$ is then related to the physical frequency by $\Omega=\sqrt{M/r_0^3}$, and from Eq.~\eqref{Omega1} (with $\Omega_1=0$), $r_1$ is given by $r_1=-F^r_1(r_0)/[3(\Omega_0 u^t_0)^{2}f_0]$. 

At first order, we solve Eq.~\eqref{tildeEFE1} using standard frequency-domain methods~\cite{Akcay:11, Wardell-Warburton:15}, choosing a value of $r_0$ and enforcing outgoing boundary conditions $R^1_\modefont{ ilm}\propto e^{i\modefont{ m}\Omega r^*}$ at infinity and ingoing conditions $R^1_\modefont{ ilm}\propto e^{-i\modefont{ m}\Omega r^*}$ at the horizon. For stationary ($\modefont{ m}=0$) modes, we enforce regularity at the boundaries. Because $R^1_\modefont{ ilm}$ carries a flux of energy and angular momentum into the BH, the BH mass and spin slowly change, becoming $M_{BH}=M+\e \delta M(\tilde t)+{\cal O}(\e^2)$, for example. To account for this, in our first-order solutions we include mass and angular momentum perturbations proportional to $\delta M(\tilde t)$ and $\delta J(\tilde t)$, respectively. At $\tilde t_0$, we may freely specify $\delta M$ and $\delta J$, and we use that freedom to set $\delta J(\tilde t_0)=0$, making the BH nonspinning through ${\cal O}(\e)$. However, we cannot specify $\frac{d\delta M}{d\tilde t}(\tilde t_0)$ and $\frac{d\delta J}{d\tilde t}(\tilde t_0)$, as they are determined by the field equations.

At second order, we specialize our demonstration to the monopole mode, noting that through the nonlinear source $\delta^2 G_\modefont{ilm}$ in Eq.~\eqref{tildeEFE2}, this one mode of $h^{2}_{\mu\nu}$ is sourced by {\em all} modes of $h^1_{\mu\nu}$. Since $\modefont{m}=0$ if $\modefont{l}=0$, the $\modefont{l}=0$ mode of $h^2_{\mu\nu}$ is independent of the fast time $\phi_p$. There are also only four nonzero components for $\modefont{ l}=0$: $\bar h^2_{tt}$, $\bar h^2_{tr}$, $\bar h^2_{rr}$, and the trace $g^{\mu\nu}\bar h^2_{\mu\nu}$ (corresponding to $\modefont{i}=1,\,2,\,3$, and $6$). The field equations for these components split cleanly into a dissipative sector (antisymmetric under reversal of $t$ and $\phi$) and a conservative sector (symmetric under that reversal).

\emph{Dissipative sector.}
For $\modefont{ l}=0$, the dissipative sector is confined to the field equations for $\bar h^2_{tr}$ (corresponding to $\modefont{ i}=2$). Equation~\eqref{tildeEFE2} has the schematic form $\partial^2_r \bar h^{2\res}_{tr} \sim S^2_{tr} +\partial_{\tilde t}\bar h^1_{tt}+\partial_{\tilde t}\bar h^1_{rr}$, where $S^2_{tr}$ includes both the $\delta^2G_{tr}$ term and the puncture terms. We can express the $\tilde t$ derivatives in terms of the slow evolution of system parameters by writing $\partial_{\tilde t} \bar h^1_{\alpha\beta}=\frac{dr_0}{d\tilde t}\partial_{r_0}\bar h^1_{\alpha\beta}+\frac{d\delta M}{d\tilde t}\partial_{\delta M}\bar h^1_{\alpha\beta}$; we can also note that the $\delta^2 G_{tr}$ term is directly proportional  to the flux of gravitational energy across a surface of constant $r$~\cite{Martel:03}. Solving the field equations hence yields $\bar h^{2}_{tr}$ as a function of $dr_0/d\tilde t$, $d\delta M/d\tilde t$, ${\cal F}_{\cal H}$ (the flux down the horizon), and ${\cal F}_\infty$ (the flux to infinity). 

The relationship between those four quantities is then dictated by the gauge condition, which has the schematic form $\partial_r \bar h^2_{tr}\sim \partial_{\tilde t}\bar h^2_{tt}$. Enforcing this condition on our solution yields conservation equations: $\frac{d}{d\tilde t}\delta M = {\cal F}_{\cal H}$, which tells us that the BH's mass grows at the rate that energy is carried into it; and the balance law $\frac{d{\cal E}_0}{d\tilde t} = -{\cal F}_{\cal H}-{\cal F}_\infty$, which tells us that the particle's orbital energy decreases at the rate that energy is carried out across the boundaries. Here $\e{\cal E}_0=m\frac{1-2M/r_0}{\sqrt{1-3M/r_0}}$ is the zeroth-order orbital energy, and $\frac{d{\cal E}_0}{d\tilde t}=\frac{d{\cal E}_0}{dr_0}\frac{dr_0}{d\tilde t}$. These conservation laws are well established~\cite{Galtsov:82}, but our derivation of them stands as the first major test of our formalism.

\emph{Conservative sector.}
For $\modefont{ l}=0$, the conservative sector comprises $\bar h^2_{tt}$, $\bar h^2_{rr}$, and the trace of $\bar h^2_{\mu\nu}$ (corresponding to $\modefont{ i}=1,3,6$). Equation~\eqref{tildeEFE2} becomes three coupled radial equations for these components, constrained by the gauge condition. Unlike in the dissipative sector, these equations involve only the ``instantaneous'' state of the system, with no $\tilde t$ derivatives appearing. We solve the coupled equations numerically, subject to the boundary conditions determined by the PM and near-horizon expansions.

As a physical output of our calculation, we compute the binary's (specific) binding energy, which we define as $E_{\rm bind} = (M_B-\Mirr-m)/\mu$. Here $M_B$ is the system's Bondi mass, $\Mirr$ is the BH's perturbed mass, and $\mu=m \Mirr/(m+\Mirr)$ is the binary's reduced mass. We take $\Mirr$ to be the BH's irreducible mass, defined from its surface area ${\cal A}$ as $\Mirr = \sqrt{\frac{\cal A}{16\pi}}$~\cite{Christodoulou:70}. For readers interested in the technical details of our definitions, we provide a discussion in the Supplemental Material~\cite{supplemental-material}.

Both $M_B$ and $M_{BH}$ are expressed as expansions in powers of $\e$: $M_B=M+\e ({\cal E}_0+\delta M)+\e^2 M^{(2)}_B+{\cal O}(\e^3)$ and $\Mirr = M + \e\delta M+\e^2 \Mirr^{(2)}+{\cal O}(\e^3)$. Given these expansions, the binding energy is
\beq\label{Ebind_Taylor}
E_{\rm bind} = \hat {\cal E}_0-1 + \e\left(\hat M^{(2)}_B - \Mhatirr^{(2)} + \hat {\cal E}_0 - 1\right) + {\cal O}(\e^2),
\eeq
where $\hat {\cal E}_0:={\cal E}_0/M$, $\hat M^{(2)}_B:=M^{(2)}_B/M$, and $\Mhatirr^{(2)}:=\Mirr^{(2)}/M$. Here all quantities are functions of $r_0/M$, making them dependent on the Schwarzschild radial coordinate and the nonphysical background mass $M$. We remove those dependences by reexpressing all quantities as functions of the physical parameter $y=(M_{BH}\Omega)^{2/3}$. From $r_0/M = (M\Omega)^{-2/3}$ and $\Omega=y^{3/2}/M_{BH}$, we have $r_0/M=\big[1+\frac{2}{3}\e\hat{\delta M}+{\cal O}(\e^2)\big]/y$, where $\hat{\delta M}:=\delta M/M$. Substituting this into Eq.~\eqref{Ebind_Taylor} yields
\beq\label{Ebind_Taylor2}
E_{\rm bind} = \hat{\cal E}_0(y)-1 + q E_{SF} + {\cal O}(q^2),
\eeq
where $q:=m/M_{BH}$, $\hat{\cal E}_0(y):=\frac{1-2y}{\sqrt{1-3y}}$, and 
\begin{align}\label{ESF}
E_{SF} &= \hat M^{(2)}_B - \Mhatirr^{(2)} +\hat{\cal E}_0(y) - 1+ \frac{\hat{\delta M}(1-6y)y}{3(1-3y)^{3/2}}.
\end{align}
We stress that $\hat M^{(2)}_B$ and $\Mhatirr^{(2)}$ are calculated from the {\em second}-order metric perturbation at infinity and on the horizon, respectively; $E_{SF}$ appears with only a linear factor $q$ in Eq.~\eqref{Ebind_Taylor2} simply because $E_{\rm bind}$ is normalized by $\mu$.

\begin{figure}[t]
\begin{center}
\includegraphics[width=\columnwidth]{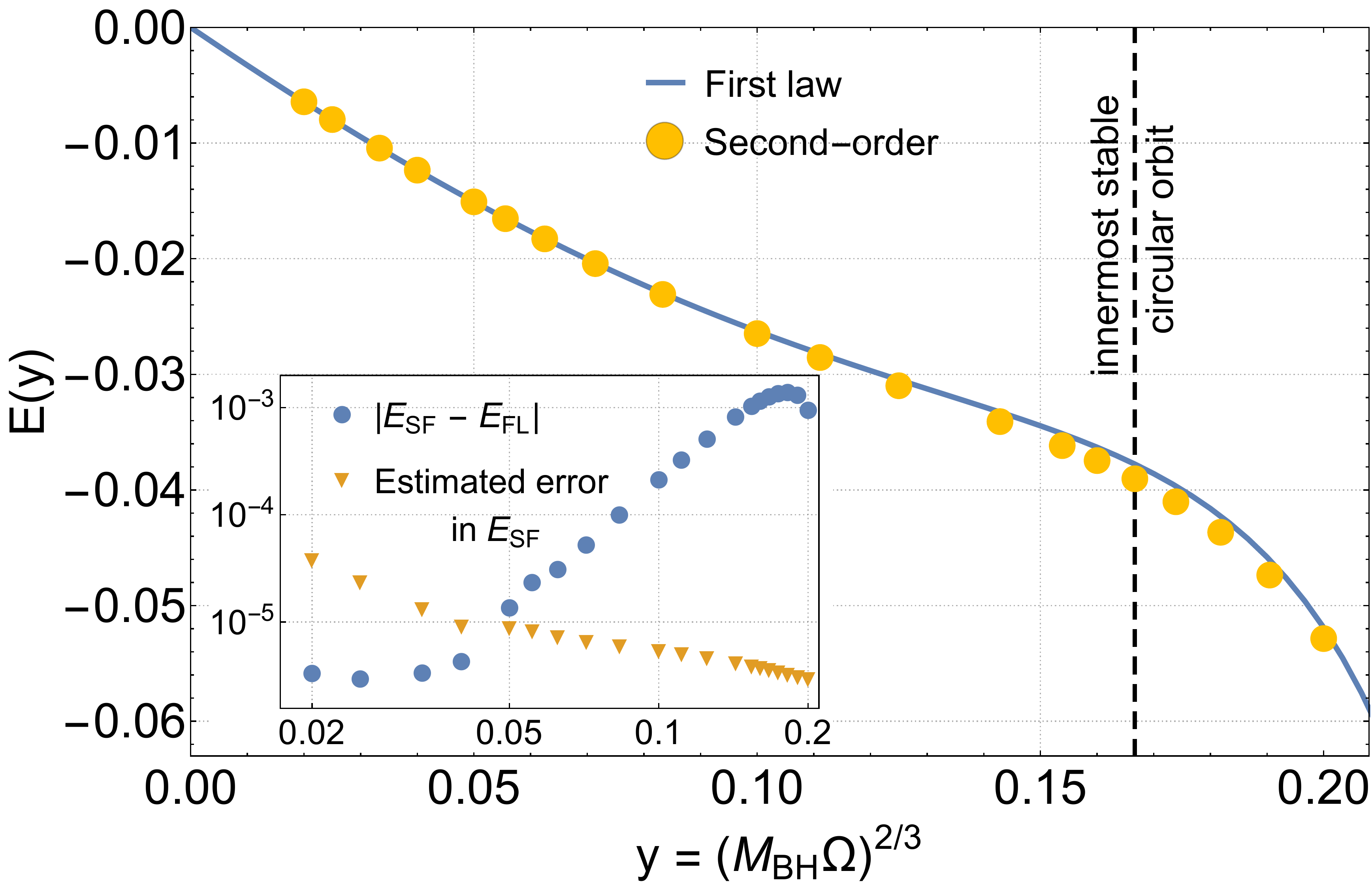}
\end{center}
\caption{\label{plot} Self-force correction to the binding energy as a function of the dimensionless parameter $y$. Increasing $y$ corresponds to decreasing orbital radius. The yellow circles show our numerical results for Eq.~\eqref{ESF}. 
The blue curve shows the mechanical binding energy~\eqref{ESF_1stLaw} derived from the first law of binary mechanics, plotted using high-accuracy numerical data from~\cite{Akcay-etal:12}. The inset shows the absolute difference between the two (blue circles) along with a conservative estimate of our absolute numerical error (orange triangles). We note that the values for $E_{SF}$ and $\hat{\mathcal{E}}_0-1 \approx - \frac{y}{2}$ are comparable, so the contribution of $E_{SF}$ to the total binding energy is suppressed by a factor of $q$ relative to the leading term.}
\end{figure}

Figure~\ref{plot} displays our numerical results for $E_{SF}$ as a function of $y$ ($\sim M/r_0$). The line marking the innermost stable circular orbit (ISCO) divides the results into two physical scenarios. Points to the left correspond to snapshots of inspiralling orbits. Points to the right correspond to snapshots of orbits that spiral {\em outward}, toward larger $r_0$, along a sequence of unstable circular orbits~\cite{Gundlach-etal:12}. The complete data is available in~\cite{supplemental-material}.

Although this marks the first direct calculation of $E_{\rm SF}$, a certain, locally defined {\em mechanical} binding energy has previously been calculated in a quite different way, based on a result from the first law of binary mechanics (FLBM)~\cite{LeTiec-etal:12b}, a powerful tool with a host of recent applications (reviewed in~\cite{Blanchet-LeTiec:17}). The FLBM relates a binary's second-order energy to the first-order Detweiler redshift $z_{SF}$~\cite{LeTiec-etal:12a}, which is a measure of proper time in the effective metric $g_{\mu\nu}+h^{\res}_{\mu\nu}$~\cite{Detweiler:08}. If we express Eqs.~(3b) and (4a) from~\cite{LeTiec-etal:12b} in terms of our variable $y$, then the FLBM's form of $E_{SF}$ is given by
\begin{align}
E^{\rm 1st\ law}_{SF} &= \frac{1}{2}z_{SF}(y)-\frac{y}{3}\frac{dz_{SF}}{dy}-1+\sqrt{1-3y} \nonumber\\
			&\quad + \frac{y}{6}\frac{5-12y}{(1-3y)^{3/2}}.\label{ESF_1stLaw}
\end{align}

Returning to Fig.~\ref{plot}, we see that $E^{\rm 1st\ law}_{SF}$ agrees fairly well with our results for $E_{SF}$, but the inset shows that the disagreement is larger than our numerical uncertainty. This is not surprising: although various formulations of the FLBM have been derived~\cite{LeTiec-etal:12a,Blanchet-etal:13,Gralla-LeTiec:13,LeTiec:15b,Fujita-etal:16,Blanchet-LeTiec:17}, none of them precisely applies to our particular scenario. Each of them is derived for a fully conservative spacetime with an exact helical Killing vector $k^\mu=(1,0,0,\Omega)$, corresponding to a binary in an {\em eternally} circular orbit. Our calculation, on the other hand, applies to an evolving system satisfying retarded boundary conditions. 

In addition to that critical difference, the version of the FLBM that is most applicable to our scenario, derived in~\cite{Fujita-etal:16}, does not define the binding energy from the system's Bondi mass. Instead, as alluded to above, it uses a local, mechanical energy. Our results suggest that this mechanical energy can approximately, but not precisely, be identified with $M_B$. A different, earlier version of the FLBM {\em does} work with an energy measured at infinity~\cite{LeTiec-etal:12a}, but it is restricted in other ways: it is derived for a system of two point particles, not a particle orbiting a large BH, and only for a nonphysical spacetime that is both helically symmetric and asymptotically flat. No exact solution to the Einstein equations satisfies those last two criteria; physically, maintaining the circular orbit would require an influx of energy from infinity, preventing asymptotic flatness. 

However, despite these limitations, NR simulations have shown that the FLBM applies remarkably well to BH binaries even in the presence of dissipation~\cite{LeTiec-etal:12b,Zimmerman-etal:16}. Our results further illuminate this, demonstrating that the FLBM applies to a good approximation through second order in perturbation theory, to snapshots of the physical, evolving spacetime of an EMRI over an orbital timescale $t\sim M$ around a constant-$\tilde t$ slice. Only in the strong-field region, for orbits near (and below) the ISCO, does the FLBM begin to substantially disagree with our full, physical result. 

The Supplemental Material contains a more thorough discussion of the comparison.

\emph{Future directions.}
The methods we have described in this letter will be the launching point for numerous calculations. In the near term, we will apply our framework to compute higher $\modefont{ lm}$ modes, from which we can extract the binary's angular momentum, the second-order Detweiler redshift~\cite{Pound:14c} and other conservative quantities, and the fluxes of energy and angular momentum carried off in gravitational waves. With all the modes in hand, we will also compute the second-order SF and hence obtain second-order-accurate inspirals and their emitted waveforms. Ultimately, we will work to extend our method to the astrophysically realistic scenario of noncircular orbits in Kerr spacetime.


\begin{acknowledgments}
{\em Acknowledgments}. We thank Leor Barack for countless helpful discussions, direct contributions to this work, and comments on a draft of this letter. We thank Alex Le Tiec for numerous informative discussions of the first law of binary mechanics and for valuable comments on this letter. We also thank Adrian Ottewill for assistance with determining analytic expressions for the even, static modes of the first-order metric perturbation. AP additionally thanks Aaron Zimmerman, Eanna Flanagan, Guillaume Faye, and Jordan Moxon for helpful discussions, and particularly Luc Blanchet for invaluable guidance in navigating the post-Minkowskian literature. Finally, we are grateful for the many illuminating conversations we have had with participants of the Capra meetings --- at University College Dublin, Caltech, Kyoto University, the Paris Observatory, the University of North Carolina at Chapel Hill, the Albert Einstein Institute in Potsdam-Golm, and the Centro Brasileiro de Pesquisas Físicas --- held since this work began in 2013.

AP acknowledges support from a Royal Society University Research Fellowship and a Natural Sciences and Engineering Research Council of Canada Postdoctoral Fellowship. AP and JM acknowledge support from the European Research Council under the European Union's Seventh Framework Programme (FP7/2007-2013)/ERC Grant No. 304978. NW gratefully acknowledges support from a Royal Society - Science Foundation Ireland University Research Fellowship, a Marie Curie International Outgoing Fellowship (PIOF-GA-2012-627781), and an Irish Research Council EMPOWER Fellowship. This material is based upon work supported by the National Science Foundation under Grant Number 1417132. B.W. was supported by Science Foundation Ireland under Grant No.~10/RFP/PHY2847, by the John Templeton Foundation New Frontiers Program under Grant No.~37426 (University of Chicago) - FP050136-B (Cornell University), and by the Irish Research Council, which is funded under the National Development Plan for Ireland.
\end{acknowledgments}

\bibliographystyle{apsrev4-1}
\bibliography{bibfile}

\end{document}


\title{Supplemental material} 
%
\maketitle

\setcounter{equation}{0}
\setcounter{figure}{0}
\setcounter{table}{0}
\makeatletter
\renewcommand{\theequation}{S\arabic{equation}}
\renewcommand{\thefigure}{S\arabic{figure}}
\renewcommand{\bibnumfmt}[1]{[S#1]}
\renewcommand{\citenumfont}[1]{S#1}

In this supplemental material, we provide (i) additional technical details on our calculation of the binding energy, (ii) a table of our numerical results, (iii) a discussion of various potential causes of disagreement with the binding energy from the FLBM, (iv) a comparison with the PN binding energy.

\section{Calculation of the binding energy}

As described in the body of the letter, we define the specific gravitational binding energy as $E_{\rm bind}=(M_B-m-M_{BH})/\mu$. Here we describe more precisely our calculations of the Bondi mass $M_B$ and BH mass $M_{BH}$.

Each of the two masses depends delicately on where it is calculated. $M_B$ is calculated on a two-dimensional cross-section of future null infinity $\scri^+$, and its value depends on which cross section we choose; this dependence arises from the fact that the system's mass changes as gravitational waves carry energy out of the system. We restrict our attention to cross sections defined by constant background retarded time $u$. Analytically matching our multiscale solution to the PM solution near $\scri^+$ then yields the expansion
\beq
M_B(u,\e) = M+\e [\delta M(\tilde u)+{\cal E}_0(\tilde u)] + \e^2 M_B^{(2)}(\tilde u)+O(\e^3), 
\eeq
where the slowly evolving coefficients are functions of $\tilde u=\e u$. This still leaves us with the freedom of choosing a value of $u$ at which to make our measurement. We can see that shifting our choice of slice by $\Delta u\sim M$ will alter $M_B^{(2)}$ by $\Delta u\frac{d}{d\tilde u}M^{(1)}_B$. To eliminate this ambiguity, we choose a preferred $u$, say $u_0$, at which the physical parameters characterizing the PM solution have the same value as those characterizing our multiscale solution at $\tilde t=\tilde t_0$. For example, $\delta M(\tilde u_0)=\delta M(\tilde t_0)$ and ${\cal E}(\tilde u_0)={\cal E}(\tilde t_0)$. $M_B^{(2)}$ is then computed from the $\modefont{l}=0$ mode of $\bar h^{2}_{\mu\nu}$ at $r\to\infty$ (after transforming our solution to a Bondi-Sachs gauge at large $r$~\cite{Madler-Winicour:16}).

Measuring $M_{BH}$ comes with analogous choices, along with new ones. As our measure of BH mass, we use the irreducible mass $\Mirr = \sqrt{\frac{\cal A}{16\pi}}$, where ${\cal A}$ is the surface area of a two-dimensional spatial cross-section ${\cal H}$ of the horizon. Since the BH is nonspinning through order $\e$, this should be a good measure of the BH's energy~\cite{Schnetter-Krishnan-Beyer:06}. In analogy with our choices at $\scri^+$, we choose ${\cal H}$ by first slicing the spacetime around the BH into surfaces of constant background advanced time $v$. Since we only have access to the metric at $\tilde t=\tilde t_0$, we do not know the location of the BH's true event horizon, which depends on the entire future history. Instead, we follow common practice in NR and take ${\cal H}$ to be the {\em apparent} horizon on each slice $v={\rm constant}$. We find the location of ${\cal H}$ analytically, as a series in $\e$, by solving $\Theta=0$, where $\Theta$ is the expansion of a null congruence passing orthogonally outward across ${\cal H}$. This then yields an expansion for $\Mirr$, given by
\beq
\Mirr(v,\e)=M+\e\delta M(\tilde v)+\e^2\Mirr^{(2)}(\tilde v)+O(\e^3), 
\eeq
where $\tilde v=\e v$. $\Mirr^{(2)}$ is given by a lengthy analytical formula with the schematic form $\sim h^{2,\modefont{l}=0}_{\mu\nu}+\partial_{\tilde v}h^{1,\modefont{l}=0}_{\mu\nu}+[(h^1_{\mu\nu})^2]^{\modefont{l}=0}$, where the fields are evaluated at $r=2M$. Finally, we choose a preferred value of $v$, say $v_0$, in the same manner we chose $u_0$. Note that the term $\propto\partial_{\tilde v}h^1_{\mu\nu}$ in $\Mirr^{(2)}$ introduces an explicitly dissipative effect into our result, proportional to the energy flux ${\cal F_H}$ into the BH. 

\begin{table}[tb]
\begin{ruledtabular}
\begin{tabular}{l r r r r r}
$1/y$ 	& \multicolumn{1}{c}{$E_{SF}$}  	& \multicolumn{1}{c}{ $E_{SF}^\text{cons. approx.}$} 	& \multicolumn{1}{c}{$E_{SF}^\text{1st law}$} 	&  \multicolumn{1}{c}{$\Delta E_{SF}$} & \multicolumn{1}{c}{$\Delta E_{SF}^\text{cons. approx.}$} \\
\midrule
$5$		& $-5.285(24 \pm 29) \times 10^{-2}$	& $-5.017(74 \pm 29) \times 10^{-2}$	& $-5.19043\times 10^{-2}$	&	$1.8\times 10^{-2}$	& $3.3\times 10^{-2}$	\\
$5.25$	& $-4.735(83 \pm 30) \times 10^{-2}$	& $-4.535(90 \pm 30) \times 10^{-2}$	& $-4.60503\times 10^{-2}$	&	$2.8\times 10^{-2}$	& $1.5\times 10^{-2}$	\\
$5.5$	& $-4.366(01 \pm 32) \times 10^{-2}$	& $-4.213(67 \pm 32) \times 10^{-2}$	& $-4.22641\times 10^{-2}$	&	$3.3\times 10^{-2}$	& $3.0\times 10^{-3}$	\\
$5.75$	& $-4.101(94 \pm 33) \times 10^{-2}$	& $-3.983(91 \pm 33) \times 10^{-2}$	& $-3.96618\times 10^{-2}$	&	$3.4\times 10^{-2}$	& $4.5\times 10^{-3}$	\\
$6$		& $-3.903(07 \pm 35) \times 10^{-2}$	& $-3.810(27 \pm 35) \times 10^{-2}$	& $-3.77662\times 10^{-2}$	&	$3.3\times 10^{-2}$	& $8.9\times 10^{-3}$	\\
$6.25$	& $-3.745(93 \pm 37) \times 10^{-2}$	& $-3.672(02 \pm 37) \times 10^{-2}$	& $-3.63081\times 10^{-2}$	&	$3.2\times 10^{-2}$	& $1.1\times 10^{-2}$	\\
$6.5$	& $-3.616(46 \pm 38) \times 10^{-2}$	& $-3.556(92 \pm 38) \times 10^{-2}$	& $-3.51297\times 10^{-2}$	&	$2.9\times 10^{-2}$	& $1.3\times 10^{-2}$	\\
$7$		& $-3.408(94 \pm 41) \times 10^{-2}$	& $-3.369(14 \pm 41) \times 10^{-2}$	& $-3.32675\times 10^{-2}$	&	$2.5\times 10^{-2}$	& $1.3\times 10^{-2}$	\\
$8$		& $-3.098(59 \pm 46) \times 10^{-2}$	& $-3.079(03 \pm 46) \times 10^{-2}$	& $-3.04741\times 10^{-2}$	&	$1.7\times 10^{-2}$	& $1.0\times 10^{-2}$	\\
$9$		& $-2.854(51 \pm 49) \times 10^{-2}$	& $-2.843(93 \pm 49) \times 10^{-2}$	& $-2.82198\times 10^{-2}$	&	$1.2\times 10^{-2}$	& $7.8\times 10^{-3}$	\\
$10$	& $-2.647(60 \pm 53) \times 10^{-2}$	& $-2.641(45 \pm 53) \times 10^{-2}$	& $-2.62628\times 10^{-2}$	&	$8.1\times 10^{-3}$	& $5.8\times 10^{-3}$	\\
$12$	& $-2.308(81 \pm 59) \times 10^{-2}$	& $-2.306(38 \pm 59) \times 10^{-2}$	& $-2.29877\times 10^{-2}$	&	$4.4\times 10^{-3}$	& $3.3\times 10^{-3}$	\\
$14$	& $-2.042(40 \pm 64) \times 10^{-2}$	& $-2.041(28 \pm 64) \times 10^{-2}$	& $-2.03717\times 10^{-2}$	&	$2.6\times 10^{-3}$	& $2.0\times 10^{-3}$	\\
$16$	& $-1.828(50 \pm 71) \times 10^{-2}$	& $-1.827(93 \pm 71) \times 10^{-2}$	& $-1.82539\times 10^{-2}$	&	$1.7\times 10^{-3}$	& $1.4\times 10^{-3}$	\\
$18$	& $-1.653(86 \pm 81) \times 10^{-2}$	& $-1.653(54 \pm 81) \times 10^{-2}$	& $-1.65151\times 10^{-2}$	&	$1.4\times 10^{-3}$	& $1.2\times 10^{-3}$	\\
$20$	& $-1.508(10 \pm 87) \times 10^{-2}$	& $-1.507(91 \pm 87) \times 10^{-2}$	& $-1.50674\times 10^{-2}$	&	$9.0\times 10^{-4}$	& $7.8\times 10^{-4}$	\\
$25$	& $-1.234(26 \pm 90) \times 10^{-2}$	& $-1.234(20 \pm 90) \times 10^{-2}$	& $-1.23383\times 10^{-2}$	&	$3.5\times 10^{-4}$	& $3.0\times 10^{-4}$	\\
$30$	& $-1.04(37 \pm 13) \times 10^{-2}$		& $-1.04(37 \pm 13) \times 10^{-2}$		& $-1.04339\times 10^{-2}$	&	$3.2\times 10^{-4}$	& $3.0\times 10^{-4}$	\\
$40$	& $-7.9(65 \pm 23) \times 10^{-3}$		& $-7.9(65 \pm 23) \times 10^{-3}$		& $-7.96242\times 10^{-3}$	&	$3.7\times 10^{-4}$	& $3.6\times 10^{-4}$	\\
$50$	& $-6.4(30 \pm 37) \times 10^{-3}$		& $-6.4(30 \pm 37) \times 10^{-3}$		& $-6.43325\times 10^{-3}$	&	$5.2\times 10^{-4}$	& $5.2\times 10^{-4}$	\\
\end{tabular}
\end{ruledtabular}
\caption{Numerical results for the binding energies. The second column shows the result of our second-order self-force calculation. Our (conservative) estimate of the numerical error is given in the brackets, e.g., $-5.285(24 \pm 29) \times 10^{-2} = (-5.28524 \pm 0.00029)\times 10^{-2}$. The numerical uncertainty in all our second-order results is dominated by discretization error. The third column shows a conservative approximant to our second-order result achieved by artificially turning off the time asymmetric terms in the boundary conditions (see the main text for details). The fourth column shows the result from the first law of binary mechanics, as computed using the high-accuracy data in \cite{Akcay-etal:12}; all digits shown in this column are accurate. The fifth and sixth columns shows the relative difference between the FLBM binding energy and the full second-order result and its conservative approximant, respectively. All the data in this table can also be found digitally in the Black Hole Perturbation Toolkit \cite{BHPToolkit}.}
\label{data}
\end{table}

\section{Comparison with the first law of binary mechanics}
 
Although there is prior numerical evidence, obtained by Zimmerman et al.~\cite{Zimmerman-etal:16}, that the FLBM applies remarkably well to inspiraling BH binaries, that numerical study also found disagreement with the FLBM at a level comparable to the difference between the 2PN and 3PN contributions. Like us, Zimmerman et al. also stressed that the comparison necessarily involves many nonunique choices. Hence, as noted in the body of the letter, we should not be surprised that our results for $E_{\rm bind}$ disagree with the binding energy obtained from the FLBM.  We now highlight several specific potential causes of disagreement.

\begin{figure}[tb]
\begin{center}
\includegraphics[width=.6\textwidth]{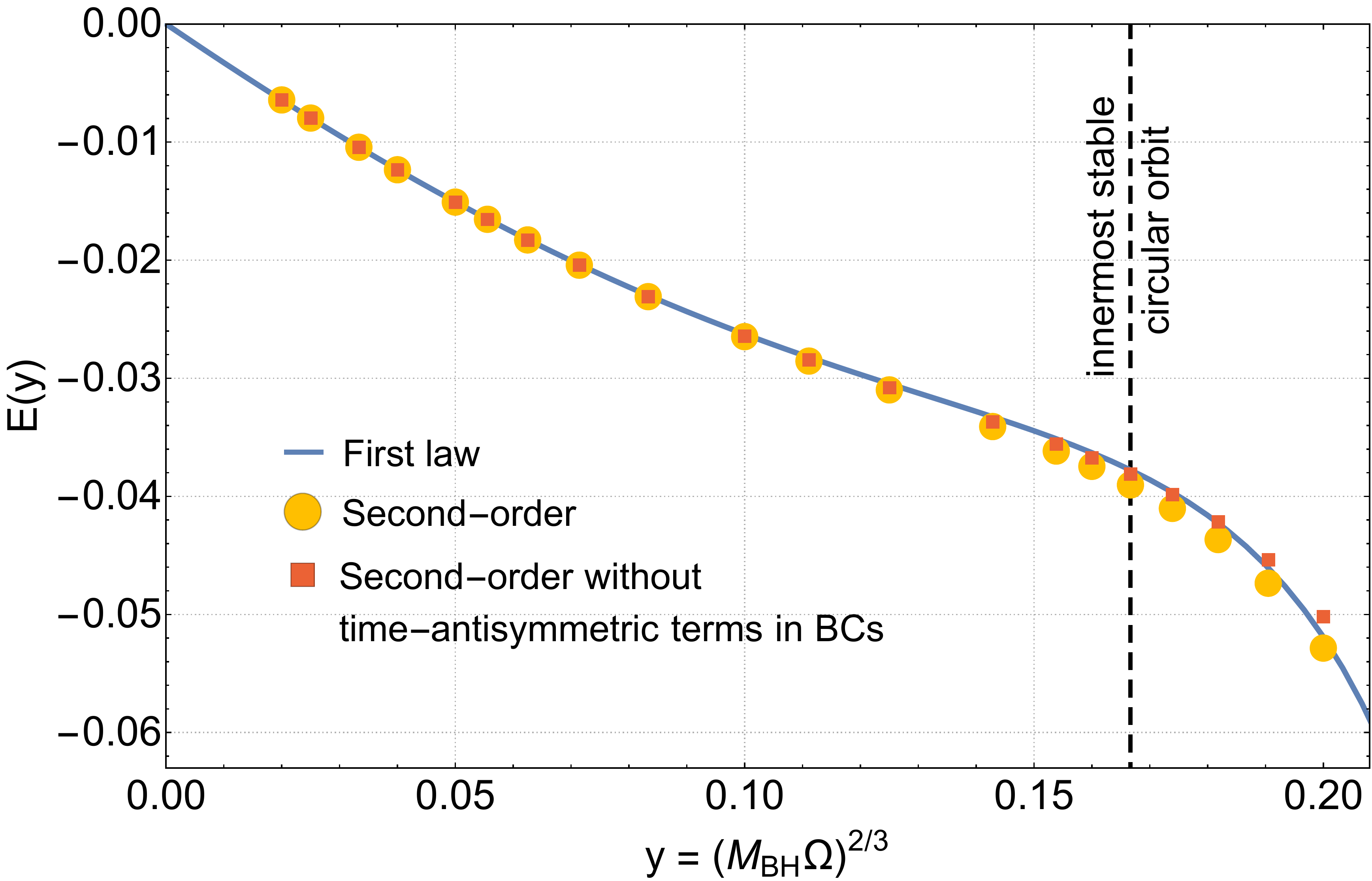}
\end{center}
\caption{\label{conservative-approximant} Effect of time-asymmetric boundary conditions. The yellow circles and blue curve are as described in Fig.~1 in the body of the letter. The red squares are obtained by artificially ``turning off'' the time-antisymmetric terms in the boundary conditions.}
\end{figure}

\begin{itemize}

\item {\bf Finite-size effects.} The version of the FLBM that works with an energy measured at infinity, rather than a local mechanical energy, is derived assuming a binary of point particles. If instead one of the bodies is a finite-sized black hole, then the method of derivation only applies in the case of corotating orbits, in which the black hole rotates with the same angular velocity as the orbiting particle~\cite{Gralla-LeTiec:13} (and even in the case of corotation, this formulation of the FLBM has only been derived at one order lower in $\e$ than is relevant here). In our scenario, with a nonrotating black hole, we may therefore expect a correction to the FLBM. There is also an inherent nonuniqueness in identifying our $M_{BH}$ with the mass of the larger body in the FLBM: as a consequence of our BH's finite and time-varying size, our value of $M_{BH}$ could change if we adopted a different slicing of the horizon, or if we  worked with the event horizon rather than the apparent horizon. We also observed above that our $M_{BH}$ includes a manifestly dissipative term, which presumably cannot appear in the FLBM. (However, this term is numerically small because it is proportional to the flux of energy down the horizon, ${\cal F}_H$, which is strongly suppressed relative to the flux to infinity~\cite{Nagar-Akcay:11}.)

\item {\bf Time asymmetric boundary conditions.} Although the $\modefont{i}=1,3,6$, $\modefont{l}=0$ field equations are time-symmetric, our {\em solution} is not, because the physical boundary conditions smuggle in explicit time-antisymmetry. The $\modefont{l}=0$ piece of the PM metric at large $r$ naturally takes the form $\e h^1_{\alpha\beta}(\e u,r,\theta,\phi)+\e^2 h^2_{\alpha\beta}(\e u, r, \theta,\phi)+\ldots$. To convert this into boundary conditions for our fields on slices of fixed $\tilde t$, we expand $h^1_{\alpha\beta}(\e u,r,\theta,\phi)$ around $h^1_{\alpha\beta}(\tilde t, r,\theta,\phi)$, giving rise to a term of the form $\e r\partial_{\tilde t}h^1_{\alpha\beta}$ in the boundary conditions for our second-order field. The same effect occurs in the boundary condition near the horizon. Such time-antisymmetric boundary conditions presumably cannot appear in the FLBM's fully time-symmetric system.

Unlike other possible causes of disagreement, this one has an easily measured impact. Figure~\ref{conservative-approximant} shows the effect of artificially turning off the time-antisymmetric terms in the boundary conditions. We see that doing so brings the binding energy closer to the FLBM's result, as one might expect. The complete data is presented in Table~I. 

\item {\bf Time-symmetric contributions of time-antisymmetric fields.}

In addition to our boundary conditions, the source in our field equations~(10) contains a subtle dependence on the time-asymmetry of our system. The second-order Einstein tensor has the schematic form $\delta^2 G_{\mu\nu}\sim \partial h^1\partial h^1+h^1\partial^2 h^1$. This means that even if we consider a time-symmetric piece of $\delta^2 G_{\mu\nu}$, that piece will contain products of time-{\em antisymmetric} first-order perturbations. It is not clear that this is compatible with the FLBM, particularly with a Hamiltonian-based formulation of the FLBM; as advocated by~\cite{Bini-Damour:16}, a Hamiltonian approach would more naturally be based on a construction using a time-symmetric Green's function at every order in perturbation theory, as opposed to our construction of the physical, retarded solution.

\item {\bf Time asymmetric hereditary contributions}

It is known from PM and PN theory that relativistic two-body dynamics can contain nonlocal-in-time, hereditary integrals that depend on the system's entire past history~\cite{Blanchet-Damour:88,Galley-etal:15}. At 4PN, this leads to a nonlocal-in-time Hamiltonian, first derived in \cite{Damour-etal:14}, which contains a certain time-symmetrized (and regularized) version of a hereditary integral. Ultimately, that integral contributes a specific local-in-time term to the FLBM's binding energy~\cite{Blanchet-LeTiec:17}. It is not clear whether such a time-symmetrized hereditary effect can appear in our calculation.

\end{itemize}

\section{Comparison with post-Newtonian binding energy}

In order to better understand the difference between the second-order self-force binding energy and the FLBM binding energy, it is instructive to compare our second-order numerical results with the analytic PN expansion of the FLBM result. Using the high-order PN expansions in Ref.~\cite{Kavanagh-etal:15}, we find that the PN expansion of Eq.~(14) is given by
\begin{align}\label{eq:ESF_1stLaw_PN}
	E^{\rm 1st\ law}_{SF} =& -\frac{y}{3}+\frac{13 y^2}{24}+\frac{35 y^3}{16}+\left(-\frac{18245}{1152}+\frac{205 \pi ^2}{192}\right) y^4 \nonumber \\
	   & + \left(\frac{719021}{11520}-\frac{448 \gamma }{15}-\frac{9037 \pi ^2}{3072}-\frac{896 \log (2)}{15}-\frac{224 \log (y)}{15}\right)y^5 + \mathcal{O}(y^6),
\end{align}
where $\gamma \simeq 0.5772$ is the Euler gamma constant. Here the $y^1$ term is the Newtonian (0PN) contribution. Successively higher powers of $y$ correspond to higher PN corrections, given above up to 4PN. We note that at these PN orders, this result for the binding energy has also been derived directly within PN theory; the PN conservative dynamics are consistent with the FLBM to at least 4PN~\cite{Blanchet-LeTiec:17}. However, this PN binding energy is again not necessarily equivalent to ours, as it is defined from a mechanical energy in the conservative, Hamiltonian dynamics. Hence, it may disagree with our results at some PN order.

\begin{figure}[tb]
	\includegraphics[width=.6\textwidth]{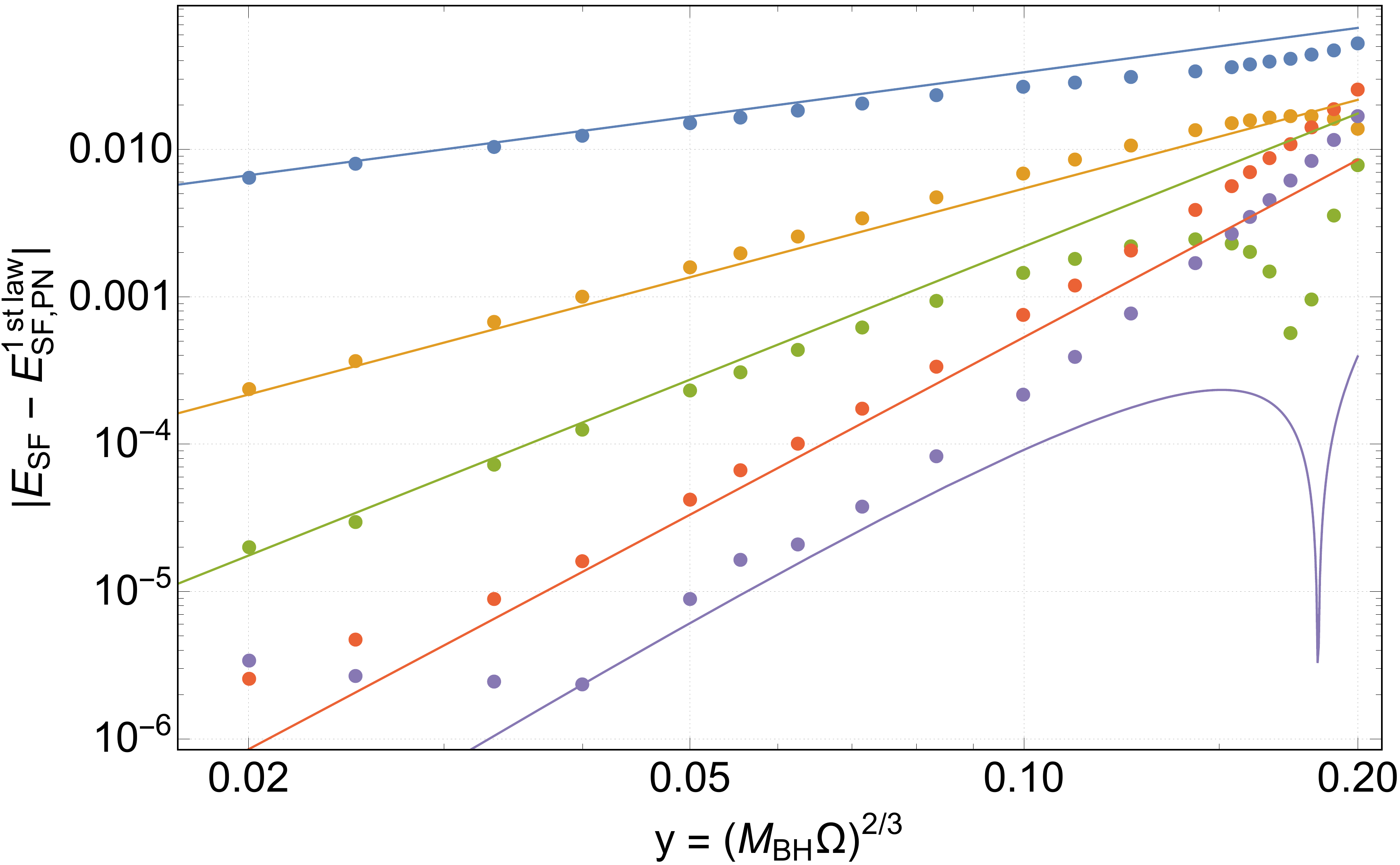}
	\caption{Comparison of our second-order binding energy results with the PN series for the FLBM binding energy. The dots show the absolute difference between our numerical results for the second-order binding energy and the successively higher PN approximations to the FLBM binding energy. Reading top to bottom along $y=0.05$, we have $E_{SF}$, $|E_{SF} - E^\text{1st law}_{SF,0PN}|$, $|E_{SF} - E^\text{1st law}_{SF,1PN}|$, etc. Along $y=0.05$ (reading top to bottom), the five solid curves show the PN series \eqref{eq:ESF_1stLaw_PN} starting at $y^n$ with $n=0,1,\dots,4$ respectively. After subtracting each PN approximation, the numerical results lie close to the next-order PN truncation of the series. This continues until the bottom dataset ($|E_{SF} - E^\text{1st law}_{SF,3PN}|$), which does not approach the 4PN result. Although on the figure the data appears to approach the bottom (4PN) curve, the data and the 4PN series are approximately the same magnitude but opposite in sign. At large radii (small $y$) numerical noise is visible in our data after subtracting many PN terms.}
	\label{fig:PN_comparison}
\end{figure}

To investigate this possibility, we compare successively higher truncations of the above series with our numerical results. This comparison is detailed in Fig.~\ref{fig:PN_comparison}. It strongly suggests that the two binding energies agree through the first four PN orders (0--3PN). Our numerical accuracy at large orbital radii is too low to confidently identify the 4PN ($y^5$) coefficient in our results, partly because such fitting is made challenging by the presence of $\log y$ terms. But the data does suggest that there is a disagreement at 4PN: our coefficient of $y^5$ appears to have the opposite sign from the PN result.

Notably, two of the effects we highlighted in the previous section arise at precisely this PN order. First, the time-antisymmetric terms in our boundary conditions turn out to contribute a term to $E_{\rm bind}$ equal to minus the energy flux to infinity, ${\cal F}_\infty$, which scales as $\sim y^5$. Second, time-symmetrized hereditary contributions to the conservative PN dynamics first enter at 4PN. While we can artificially turn off our time-antisymmetric boundary term, doing so does not bring our results into agreement. This indicates that the disagreement is very likely related to the time-symmetrized hereditary effects. Because those effects are subtler and more difficult to adjust for, we leave deeper investigation of this issue to future work.

\bibliography{bibfile}